\documentclass[a4paper,12pt]{article}
\usepackage{epsfig}
\usepackage{subfig}
\usepackage{amssymb}
\begin{document}
\thispagestyle{empty}

\newcommand{\etal}  {{\it{et al.}}}  
\def\Journal#1#2#3#4{{#1} {\bf #2}, #3 (#4)}
\def\PRD{Phys.\ Rev.\ D}
\def\NIMA{Nucl.\ Instrum.\ Methods A}
\def\PRL{Phys.\ Rev.\ Lett.\ }
\def\PLB{Phys.\ Lett.\ B}
\def\EPJ{Eur.\ Phys.\ J}
\def\IEEETNS{IEEE Trans.\ Nucl.\ Sci.\ }
\def\CPCD{Comput.\ Phys.\ Commun.\ }

\smallskip

\bigskip
\bigskip

{\Large\bf
\begin{center}
Towards the Bose symmetry violation issue
\end{center}
}
\vspace{1 cm}

\begin{center}
{\large G.A. Kozlov  }
\end{center}
\begin{center}
\noindent
 { Bogolyubov Laboratory of Theoretical Physics\\
 Joint Institute for Nuclear Research,\\
 Joliot Curie st., 6, Dubna, Moscow region, 141980 Russia  }
\end{center}

 \begin{abstract}
 \noindent
 {We study the Bose symmetry violation through the decays of heavy vector bosons at high energies. In particular, the decay of a $Z^{\prime}$-boson into two photons where one of the photons is the vector  
unparticle  in the scale invariant sector is considered as a sample. We find out that the Bose symmetry might be violated in the nearly conformal sector at high energy frontier.
 This may be useful in phenomenological application to the CERN LHC experiments for new physics search.  }


\end {abstract}
PACS numbers: 11.30.Qc, 11.25.Hf, 14.70.Pw




\bigskip

{\it Introduction.-}
Although there is  a bright success of the Standard Model (SM) of the interaction between elementary particles, most of efforts are actually concentrated on discussion of violation of the SM at extreme energies. Recently, the question of possible observation of a small
violation of Bose statistics at the TeV scale has been discussed in [1], where, in particular, the review of significant theoretical and
experimental efforts to motivate and find tiny departure from the connection between spin and statistics, has been done.

In accordance with the Landau [2] and Yang [3] papers, the decay of massive neutral gauge bosons into two photons are forbidden. The 3-point function relevant to the decay above mentioned has the following properties: $G^{\rho\nu\mu} (k_{1}, k_{2}) = + G^{\rho\mu\nu} (k_{2}, k_{1})$, where $k_{1}, k_{2}$ are four-momenta of photons.
The general principles of rotation, gauge invariance, the Bose statistics in order to derive certain selection rules for decays of a parent particle into two photons are used. If one of the photons becomes off-shell, the Bose symmetry is violated, and the amplitude of the process is finite.\\

The Bose symmetry violation, if any, essentially depends on the energy. The authors [1] pointed out that the search for the decay $Z^{\prime}\rightarrow\gamma\gamma$ can be effectively used to probe this violation at the LHC, if the photons do not obey Bose statistics.
The discovery of $Z^{\prime}$ or other heavy extra neutral gauge bosons with masses of the order $\sim O $(1 TeV) which would apper as resonances or even with continuous mass will give the most powerful support to the models based on the scale (or conformal) invariance as a fundamental symmetry in Nature. 

The main subject of this paper is the generalization of the Landau-Yang theorem to the case when the Bose symmetry may be violated at high energies. A very special expectation is that the model may be relevant for the description of new TeV scale physics expected to appear at the LHC.


{\it Unparticles and dilatons.-} In 1982, Banks and Zaks [4] investigated the special properties of matter with non-trivial scale invariance in infra-red (IR) region. This new kind of stuff, called as the unparticles [5], does not possess those quantum numbers which are known in the SM, in particular, has no definite mass. 
The renormalizable interactions between the SM fields
and the fields of yet hidden conformal sector could be realized by means of explore the hidden energy at high energy collisions and/or associated with the registration of non-integer number $d$ of
invisible particles. 
We suppose the existence of the unparticle stuff in the scale invariant sector of elementary particle spectrum with scale dimension $d$ of the unparticle operator. Note, that in an unitary theory fields with $d <1$ are not allowed [6]. 
The hidden vector sector at  very high energy is close to  the fundamental  ultra-violet (UV) scale $M$ at which all the particles are massless and it is far beyond the reach of the LHC. However, the presence of such a sector can affect the low energy physics [7-11]. 

It is known (see, e.g., [12, 13]) that at high enough energies, the conformal theory might be strongly coupled. Flowing down to low energies the theory might include a light scalar particle (dilaton) the properties of which are similar to those of a light $\sim O$ (100  GeV) scalar Higgs-boson in SM. The scale invariance of the strong dynamics is spontaneously broken at a scale $f$. If $f \geq v = 246 $ GeV, the strong dynamics is also responsible for the electroweak symmetry breaking (EWSB). The dilaton can appear as the Goldstone boson associated with spontaneously broken scale symmetry. It is assumed that conformal invariance is breaking by the sector containing the small parameter providing the dilaton would be lightest state compared to other resonances with masses $\sim 4\pi f$.  

The ATLAS [14]  and CMS [15] data on $\gamma\gamma$ event rates were observed the signal strength about factor 2 larger than that expected from the SM Higgs decay. This exhibits the more strong coupling of the scalar particle with massless gauge bosons beyond the SM. One of the candidate having such the properties is the dilaton which has the enhancement [13,11,16] of $\gamma\gamma$ and $gg$ couplings compared to the SM Higgs boson. 
One of the main messages is that  at some energy region $\tilde\Lambda < Q < \Lambda $  the (hidden) conformal sector may flow to IR fixed point and involve the coupling of  the dilaton  to an otherwise hidden sector (unparticle) by some mediator (ultra heavy field) of mass $M$:  $L_{UV} \sim \epsilon _{k}\,O_{\sigma}\,O_{U}$, where $\epsilon_{k} =  \Lambda ^{d_{BZ} - d}/M^{k}$,  $k = [d_{UV} - (4-n)] >0$;  $d_{BZ}$, $d_{UV}$ and $n$ mean the scale dimensions of the Banks-Zaks  sector operator, the operators in the UV and in broken conformal symmetry sector, respectively; $\Lambda  < M$, $\tilde\Lambda$ is the energy scale below which there is the SM fields sector; $O_{\sigma}$ and $O_{U}$ are the operators of the $\sigma$-dilaton and $U$- unparticle stuff, respectively. If $M$ is large enough, the unparticle stuff just doesn't couple strongly to dilaton fields to have been seen. However, it could show up at higher energies $\sim\epsilon_{k} $. Evolve down to lower energies, $Q < \Lambda$,  $L_{UV}\rightarrow L_{IR} \sim \epsilon_{k}\,O_{\sigma}\,O_{IR}$. The violaion of the SM could be predicted by the processes in IR. As far as $\epsilon_{k}\rightarrow 0$ (i.e., $\Lambda\rightarrow\tilde\Lambda\sim f\sim v$), $L_{IR}\rightarrow L_{SM}$. 

For the experimentalists the question arises: how does an unparticle stuff begin to show up as the energy of an experiment increases? The standard physical process is provided by the following amplitude
$$ {\vert\langle S P_{state_{out}}\vert\epsilon_{k}\,O_{\sigma}\,O_{U}\vert S P_{state_{in}}\rangle\vert}^{2} = \epsilon_{k}^{2}{\vert\langle S P_{state_{out}}\vert O_{\sigma}\vert S P_{state_{in}}\rangle\langle U\vert O_{IR}\vert 0\rangle\vert }^{2} $$
as $f\rightarrow v$.
The physical results in production of unparticle stuff would be seen in missing energy and momemtum on the level $\sim  O(\epsilon_{k}^{2})$.


 A nearly conformal theory (that works at $\tilde\Lambda < Q < \Lambda $) may have a special case $b^{heavy}_{0} + b^{light}_{0} = 0$ for the first coefficients of $\beta$-function applied to heavy and light fermion degrees of freedom. We are at the stage when massless particles and unparticle stuff appear in the sector of conformal breaking scale $\sim f$.



{\it Anomalous triple coupling.-} The unparticle production at hadron colliders will be a signal that the scale where conformal invariance becomes important
for particle physics is as low as a few TeV.
One can suggest  that, somehow, a series of new reactions that involve unparticle stuff in an essential way  would be at LHC  energies. 
This can be done through the study of $ pp \rightarrow Z^{\prime}\rightarrow \gamma + \gamma_{U}$, where $\gamma_{U}$ is identified with $U$ vector unparticle in the nearly conformal sector. Actually, $\gamma_{U}$ is the standard photon when $f = v$.

Before to use the concrete model, we have to make the following retreat.
First of all, we go to the extension of the Landau - Yang theorem [2,3] for the decay of a vector
particle into two vector states. The direct interaction between  $Z^{\prime}$ - boson
and $\gamma_{U}$, accompanied by a photon, does not exist. To the lowest order of the coupling constant $g$,  the contribution given by  $g^3$ in the decay
$Z^{\prime}\rightarrow \gamma \gamma_{U}$ is provided mainly
by heavy quarks in the loop. It is worth to remember the  calculation of the anomaly triangle diagram $ZZ\gamma$ [17], where the anomaly contribution result contains
two parts, one of which has no the dependence of the mass $m_{f}$ of (intermediate) charged fermions in the loop, while the second part is proportional to  $m^{2}_{f}$. 
The amplitude of the decay $Z^{\prime}\rightarrow \gamma\gamma_{U}$ is induced by the anomaly effect.
The contribution from light quarks with the mass  $m_{q}$ is suppressed as
 $m^{2}_{q}/m^{2}_{Z^{\prime}}\sim 10^{-8} - 10^{-6}$, where $m_{Z^{\prime}}$ is the mass of $Z^{\prime}$ - boson.
Despite the decay  $Z^{\prime}\rightarrow \gamma \gamma_{U}$ is the rare process, there is a special attention to the sensitivity of this decay to top-quark and even to quarks of fourth generation.
Since the photon has the only vector nature of interaction with the quarks, the possible types
of interaction $Z^{\prime} - \gamma_{U} - \gamma$ would be either $ V - A - V$ or  $ A - V - V$, where $V (A)$ means the vector (axial-vector) interaction.

{\it Constrains in the  hidden sector.-} In order to estimate the experimental constraints in hidden (conformal) sector one can start with the effective coupling between the $\sigma$-dilaton and unparticle stuff, given by operator $O_{UV}$ in UV:
 \begin{equation}
\label{eq1}
L_{\sigma UV}\sim M^{2-d_{UV}}\,{\vert \sigma\vert}^{2}\,O_{UV}.
\end{equation}
Below the strong scale $\Lambda$, the couplings (\ref{eq1}) flow to 
\begin{equation}
\label{eq2}
L_{\sigma IR}\sim C\,\frac{\Lambda ^{d_{UV} - d}}{M^{d_{UV}-2}}\,{\vert \sigma\vert}^{2}\,O_{IR},
\end{equation}
where $d$ is the dimension of the operator $O_{IR}$ in IR. Near the transition of the conformal symmetry breaking one can expect ${\vert \sigma\vert}^{2}\rightarrow f^{2}$, $O_{IR}\rightarrow \tilde\Lambda ^{d}$. The coupling  (\ref{eq2}) transforms as 
\begin{equation}
\label{eq3}
L_{\sigma IR}\sim C\,\frac{\Lambda ^{d_{UV} - d}}{M^{d_{UV}-2}}\, f^{2}\,\tilde\Lambda ^{d}
\end{equation}
and the bound on $\tilde\Lambda$ is 
\begin{equation}                     
\label{eq4}
\tilde\Lambda =\left (\frac{\Lambda^{d_{UV}-d}}{M^{d_{UV} - 2}}\,f^{2}\right )^{1/(4-d)} < \Lambda.
\end{equation}

{\it NP observables.-} Any New Physics (NP) observables involving operators in (\ref{eq2}) 
will be given by the quantity
\begin{equation}
\label{eq4}
\hat o = \epsilon_{k}^{2}\,Q^{2[d-(4-n)]}
\end{equation}
with the actual energy of the experiment $Q\sim O(\sqrt {s})$, $\sqrt {s}$ is the center-of-mass system energy, $\tilde\Lambda < Q$, $n$ is the dimension of the SM operator. This means that NP with unparticles will only be visible at the experiment where
\begin{equation}
\label{eq5}
Q > \hat o^{1/(2n)}\,M\,\left (f/M\right )^{2/n}.
\end{equation}
The important feature in (\ref{eq5}) is that of no the dependence of the dimensions of UV and unparticle stuff operators. 

{\it Lagrangian density-} At energies $Q\sim \Lambda $ the Lagrangian density (LD) is $L= L_{1} + L_{2}$, where 

\begin{equation}
\label{eq6}
L_{1} = \sum_{k=1}^{N} \left [-\frac{1}{4}B^{\mu\nu,k}\,B_{\mu\nu,k} + \frac {m^{2}_{k}}{2}\left (B_{\mu,k} - \partial \sigma_{k}\right )^{2}\right ], 
\end{equation}
\begin{equation}
\label{eq7}
L_{2}  =  g_{Z^{\prime}}\sum_{q} \bar q (v^{\prime}_{q}\,\gamma^{\mu} -
a^{\prime}_{q}\,\gamma^{\mu}\,\gamma_{5}) q \,Z^{\prime}_{\mu} + 
\frac{1}{\Lambda^{d-1}}\sum_{q} \bar q (c_{v}\,\gamma^{\mu} -  a_{v}\,\gamma^{\mu}\gamma_{5}) q\, \gamma_{U_{\mu}}.
\end{equation}
In the gauge invariant LD (\ref{eq6})  ($B_{\mu,k}\rightarrow B_{\mu,k} + \partial_{\mu}\alpha_{k}$) the dilaton field $\sigma$  serves as the conformal compensator which under gauge transformation shifts as the Goldstone boson, that is $\sigma_{k}\rightarrow \sigma_{k} + \alpha_{k}$; $B_{\mu\nu,k} = \partial_{\mu} A_{\nu,k} - \partial_{\nu} A_{\mu,k}$, $m_{k}$ is the mass of $k$th vector field,  $g_{Z^{\prime}} = (\sqrt {5b/3}\,s_{W}\,g_{Z})$ is the gauge constant of $U^{\prime}(1)$ group
(the coupling constant of $Z^{\prime}$ with a quark $q$) with the group factor $\sqrt {5/3}$,
$b\sim O(1)$, $g_{Z}=g/c_{W}$; $s_{W}(c_{W})= \sin\theta_{W} (\cos\theta_{W})$, $\theta_{W}$ is the angle of weak
interactions (often called as Weinberg angle); $v^{\prime}_{q}$ and $a^{\prime}_{q}$ are generalized vector and
the axial-vector  $U^{\prime}(1)$ -charges, respectively. These latter charges are dependent on both (joint) gauge group  and the Higgs representation which is responsible for the breaking of initial gauge group to the SM; $c_{v}$ and $a_{v}$ are unknown vector and axial-vector couplings.
Actually, the second term in $L_{2}$ (\ref{eq7}) is identical to the first one up to the factor
$\Lambda^{1-d}$.

Following the paper [18]  we introduce the new vector field $B_{\mu} = \sum_{k=1}^{N} c_{k}\,B_{\mu,k}$, the propagator of which in transverse gauge in the limit $N\rightarrow \infty$ is
\begin{equation}
\label{eq8}
D_{\mu\nu} (q) = \left (g_{\mu\nu} - \frac{q_{\mu}\,q_{\nu}}{q^{2}}\right )\int_{0}^{\infty}\frac{\rho (t)}{q^{2} - t + i\epsilon} dt,
\end{equation}
where the spectral density is $\rho (t) = \vert c^{2}_{k}\vert\delta (t - m^{2}_{k})$.

The two-point correlation function of $\gamma_{U}$ unparticle operator is 
\begin{equation}
\label{eq9}
\Pi_{\mu\nu} (q) = \int d^{4} x\langle 0\vert T \gamma_{U_{\mu}}(x) \gamma_{U_{\nu}}(0)\vert 0\rangle = 
\left (- g_{\mu\nu} + \xi\,\frac{q_{\mu}\,q_{\nu}}{q^{2}}\right )\int_{0}^{\infty}\,\rho_{0}(t)\frac{i}{q^{2} - t + i\epsilon} \frac{dt}{2\,\pi}, 
\end{equation}
where $\rho_{0}(t) = A_{d}\,t^{d-2}$, $\xi$ is unknown parameter, and $d$- dependent phase-space factor is [5]
$$A_{d} =\frac{16\,\pi^{5/2}}{(2\,\pi)^{2d}}\, \frac{\Gamma(d+1/2)}{\Gamma(d-1)\,\Gamma(2d)}.$$
  The spectrum of $\gamma_{U}$ operator is continuous and (\ref{eq9}) looks like:
\begin{equation}
\label{eq10}
\Pi_{\mu\nu} (q) =  \left (- g_{\mu\nu} + \xi\,\frac{q_{\mu}\,q_{\nu}}{q^{2}}\right )\frac{i\,A_{d}}{2\,\sin (\pi\,d)}\,\left [-(t-m_{g}^{2}) - i\epsilon\right ]^{d-2}.
\end{equation}
The mass gap $m_{g}$ in (\ref{eq10})  is introduced to eliminate IR divergence in the propagator. If $m_{g} > m_{\bar\psi} + m_{\psi}$, the unparticle can decay into SM particles, e.g., the fermion $\psi$  and the antifermion $\bar\psi$ with the masses $m_{\psi}$ and $m_{\bar\psi}$, respectively (production threshold). The mass relation $m_{g} < m_{\bar\psi} + m_{\psi}$ corresponds to a stable unparticle. The location of the unparticle resonance with the mass $m_{\gamma_{U}}$ and the width $\Gamma_{\gamma_{U}}$  is given by the propagator at $t = m^{2}_{\gamma_{U}} - i\,m_{\gamma_{U}}\, \Gamma_{\gamma_{U}}$. For the particular case channel $\gamma_{U}\rightarrow \bar\psi\psi$, if $m^{2}_{\gamma_{U}} > (m_{\bar\psi} + m_{\psi})^{2}$, the unparticle should be detected in the corresponding cross-section through a peak in the invariant mass distribution of the final state. On the contrary, if 
 $m^{2}_{\gamma_{U}} < (m_{\bar\psi} + m_{\psi})^{2}$, then the only indirect detection of the unparticle should be by an access of events with respect to the corresponding SM cross-section. 
The function $\rho_{0} (t)$ in (\ref{eq9}) may also be given in the spectral form
\begin{equation}
\label{eq11}
\rho_{0} (t) = 2\,\pi\,\sum_{\lambda} \delta (t- m^{2}_{\lambda}) {\vert \langle 0\vert\gamma_{U_{\mu}} (0)\vert\lambda\rangle \vert}^{2}.
\end{equation}
The sum in (\ref{eq11}) is in fact an integral and the spectrum of $\gamma_{U{\mu}}$ is continuous. Now, we introduce the vector state $\gamma_{U}$ as the unparticle tower in the form $\gamma_{U{\mu}} (x) = \sum _{k} c_{k}\,\gamma_{U_{\mu,k}} (x)$ with the normalization condition $\sum_{\lambda}{\vert \langle 0\vert\gamma_{U_{\mu}} (0)\vert\lambda\rangle\vert}^{2} =1$, ${\vert 0\langle \gamma_{U_{\mu}} (0)\vert\lambda_{k}\vert}^{2} =c^{2}_{k}$, $\langle 0\vert \gamma_{U_{\mu}} (0)\vert\lambda_{k}\rangle = \epsilon_{\mu}\,c_{k}$, $\epsilon_{\mu}$ is the polarization of $\gamma_{U_{\mu}}$. Using this tower-like scheme, the spectral function (\ref{eq11}) transforms as 
\begin{equation}
\label{eq12}
\rho_{0} (t) = 2\,\pi\,\sum_{k=1}^{\infty} \delta (t- m^{2}_{k})\,c^{2}_{k}.
\end{equation}
The discrete tower of states with the spacing is controlled by small parameter $\Delta$, in particular,  $m^{2}_{k} = N^{-1}\,k\,\Delta^{2} $, as $\Delta\rightarrow 0$.

Comparing $\rho_{0} (t)$ in (\ref{eq9}) and (\ref{eq12}), we find
\begin{equation}
\label{eq13}
c^{2}_{k} = \frac{1}{2\,\pi}\,A_{d}\,\left (m^{2}_{k}\right )^{d-2}\,\Delta^{2}.
\end{equation}
Thus, we have $N$ massive vector fields with masses $m_{k}$. The vector unparticle is stable in the classical conformal valley, $\Delta = 0$. However, in the nearly conformal case, when $\Delta$ is non-equal to 0, e.g., at the breaking scale of conformal invariance $\sim \tilde\Lambda$, the lifetime is finite $\sim 1/c_{k}^{2}\sim 1/\Delta^{2}$, and the unparticle can decay into SM particles, e.g., fermions. In particular, the decay width of $\lambda_{k}$ unparticle into a lepton $l$ and an antilepton $\bar l$ is 
\begin{equation}
\label{eq14}
\Gamma (\lambda_{k}\rightarrow l \bar l) = \frac{c^{2}_{A}}{8\,\pi\, \Lambda ^{2(d-1)}}\,c^{2}_{k}\,m_{k} ,
\end{equation}
where $c_{A}$ is the axial-vector constant in the vertex $c_{A}\,\Lambda ^{1-d}\,\bar l \gamma_{\mu}\gamma_{5}\, l\,\lambda^{\mu}_{k}$. Assuming that $\Lambda\sim m_{Z^{\prime}}$, $m_{k}=\delta\,m_{Z^{\prime}}$ with $0< \delta < 1$, one can estimate the lifetime of $\lambda_{k}$ as
\begin{equation}
\label{eq15}
\tau_{\lambda_{k}}\sim \frac{(4\,\pi)^{2}}{c^{2}_{A}\,A_{d}}\,\frac{\delta^{3-2d}}{\Delta^{2}}\,m_{Z^\prime},
\end{equation}
where $d\geq 3$ (the unitarity condition for primary, gauge invariant vector unparticle operator [6] is assumed). We find that the invariant lepton mass spectrum does not contain the peak in the final state, but will have the monotonuous distribution. The $\Delta$ - mass resolution can be less than that of the experimental resolution.

{\it Branching ratio-} The NP observable $\hat o_{\gamma\gamma}$ in the decay $Z^{\prime}\rightarrow \gamma\gamma$ is associated with the branching ratio 
\begin{equation}
\label{eq16}
\hat o_{\gamma\gamma} = Br (Z^{\prime}\rightarrow\gamma\gamma) = 
\frac{\Gamma (Z^{\prime}\rightarrow\gamma\gamma)}{\Gamma_{Z^{\prime}}}
\end{equation}
and the upper limit on $ \hat o_{\gamma\gamma}$ is
\begin{equation}
\label{eq17}
\hat o_{\gamma\gamma}  < \left (\frac {Q^{n}}{M^{n-2}\,f^{2}}\right )^{2}.
\end{equation}
Using the expectation value $\Gamma (Z^{\prime}\rightarrow\gamma\gamma) < 1.4 \,m_{Z^{\prime}} [TeV]$ MeV of the upper limit on the partial width of the decay $Z^{\prime}\rightarrow\gamma\gamma$ in the $Z^{\prime}$-mass scale below 5 TeV [1]  we obtain 
the upper limit $ \hat o_{\gamma\gamma} < 10^{-4}$, where the total decay width of $Z^{\prime}$ boson $\Gamma_{Z^{\prime}} = a m_{Z^{\prime}} $ has been used with $a= 0.03$ as a maximal value in the ratio $ \Gamma_{Z^{\prime}}/m_{Z^{\prime}} $ among the Grand Unification Theory inspired models [19].
If the breaking scale $f$ is close to $v$, the UV scale $M\sim 1600$ TeV which is rather heavy for the messenger between the hidden sector and the SM world. This scenario would be difficult in realization in any experiment with TeV's energy scale. On the other hand, the model we have developed in this paper with $f = 1 $ TeV yields the value for the mass $M\sim $ 100 TeV of the messenger  between the unparticle stuff and the dilaton field which has the perspectives to be explored in high energy machines. 

We have already found [11] the branching ratio of the $Z^{\prime}$ - decay into photon and the vector unparticle in the analytic form
\begin{equation}
\label{eq18}
Br^{TH} (Z^{\prime}\rightarrow\gamma\gamma_{U}) = F_{d}\,\left (\frac{m_{Z^{\prime}}^{2}}{\Lambda^{2}}\right )^{d-1}
\end{equation}
for the case, when the masses $m_{q}$ of quarks (in the loop) are less than that of $m_{Z^{\prime}}$, $m_{q} < 0.5\,m_{Z^{\prime}}$. Here, 
\begin{equation}
\label{eq19}
F_{d} = \frac{5\,A_{d}}{3\,a}\left (\frac{1}{9}+\pi^{2}\right )\left (\frac{s_{W}}{c_{W}}\right )^{2}\left (\frac{\alpha}{2\,\pi^{2}}\right )^{2}\left [\frac{1}{d (d+2)} + \frac{2}{(d+1)(d+2)}\right ].
\end{equation}

If we assume that at high energies the Bose symmetry is violated, then  $Z^{\prime}\rightarrow\gamma\gamma$ is $\Gamma (Z^{\prime}\rightarrow\gamma\gamma) = g_{B}^{2}\,m_{Z^{\prime}}$, where $g_{B}$ is the parameter characterizing the degree of Bose statistics violation. This parameter can be estimated from the experiment. We can give the answer for $g_{B}$ as the $d$-dependent function within  the $Z^{\prime}$ mass in nearly conformal sector:
\begin{equation}
\label{eq20}
g_{B}^{2} = a\, F_{d} \,\left (\frac{m_{Z^{\prime}}^{2}}{\Lambda^{2}}\right )^{d-1}.
\end{equation}
Obviously, $g_{B} = 0$ if $d=1$ in the case when the vector unparticle is given by non-primary operator through the derivative of the dilaton field, $\partial_{\mu}\sigma$.
The photon energy $E_{\gamma} = m_{Z^{\prime}} (1- z_{U})/2$ in the limit $z_{U} =P_{\gamma_{U}}^{2}/m_{Z^{\prime}}^{2}\rightarrow 0$ gets its finite value for the momentum $P_{\gamma_{U}}$ of the $\gamma_{U}$-unparticle.

{\it How to search the Bose symmetry violation?-} The experimental signature of the $Z^{\prime}\rightarrow\gamma\gamma_U$ consists of a peaks for each value of $k$ 
(see (\ref{eq12}) - (\ref{eq15})) which in the limit $\Delta\rightarrow\infty$ merge into the continuum distribution. This is the opposite to the signature of $Z^{\prime}\rightarrow\gamma\gamma$ decay, where the final state consists of two high transverse momentum ($p_{T}$) photons, $p_{T} \simeq m_{Z^{\prime}}/2$, and there will be expected a peak in the invariant mass distribution of two photons over the continuum background. 

{\it Conclusion-} To conclude, we considered a model with breaking of the conformal invariance in order to clarify the possibility of Bose statistics violation at high energies. To this, the transition of the high mass $Z^{\prime}$ boson into photon and a vector unparticle has been used as a sample. 
The parameter $g_{B}$ responsible for Bose symmetry violation has a rising behaviour with increasing of the non-integer number $d$, as well as the mass of $Z^{\prime}$ boson. 
We find out that the Bose symmetry might be violated in the nearly conformal sector at high energy frontier $\sim\Lambda$. However, the Bose statistics will be restored at energies below $\tilde\Lambda$, where $f = v$.

\end{document}